\documentclass[useAMS,usenatbib]{mn2e}
\input{epsf}
\usepackage{amssymb}
\usepackage{natbib}
\usepackage{color}
\usepackage{journals}

\newbox\grsign \setbox\grsign=\hbox{$>$} \newdimen\grdimen \grdimen=\ht\grsign
\newbox\simlessbox \newbox\simgreatbox
\setbox\simgreatbox=\hbox{\raise.5ex\hbox{$>$}\llap
     {\lower.5ex\hbox{$\sim$}}}\ht1=\grdimen\dp1=0pt
\setbox\simlessbox=\hbox{\raise.5ex\hbox{$<$}\llap 
     {\lower.5ex\hbox{$\sim$}}}\ht2=\grdimen\dp2=0pt

\newcommand{\hMpc}{{\ifmmode{h^{-1}{\rm Mpc}}\else{$h^{-1}$Mpc }\fi}}
\newcommand{\hGpc}{{\ifmmode{h^{-1}{\rm Gpc}}\else{$h^{-1}$Gpc }\fi}}
\newcommand{\hkpc}{{\ifmmode{h^{-1}{\rm kpc}}\else{$h^{-1}$kpc }\fi}}
\newcommand{\hMsun}{{\ifmmode{h^{-1}{\rm {M_{\odot}}}}\else{$h^{-1}{\rm{M_{\odot}}}$}\fi}}
\newcommand{\Msun}{{\ifmmode{{\rm {M_{\odot}}}}\else{${\rm{M_{\odot}}}$}\fi}}

\title[Cosmic variance of the local Hubble flow]{Cosmic variance of the local Hubble flow in large-scale cosmological simulations}
\author[R. Wojtak et al.]{Rados{\l}aw Wojtak$^{1}$, Alexander Knebe$^{2}$, William A. Watson$^{3}$, Ilian T. Iliev$^{3}$, 
\newauthor Steffen He{\ss}$^{4}$, David Rapetti$^{1}$, Gustavo Yepes,$^{2}$ Stefan Gottl\"ober$^{4}$
\\ 
$^1$Dark Cosmology Centre, Niels Bohr Institute, University of Copenhagen, Juliane Maries Vej 30, DK-2100 Copenhagen \O, 
Denmark\\
$^2$Departamento de F\'isica Te\'orica, Modulo C-15, Facultad de Ciencias, Universidad Aut\'onoma de Madrid, 28049 Cantoblanco, Madrid, Spain\\
$^3$Astronomy Centre, Department of Physics \& Astronomy, Pevensey II Building, University of Sussex, Falmer, Brighton, BN1 9QH, UK \\
$^4$Leibniz-Institute f\"ur Astrophysik Potsdam (AIP), An der Sternwarte 16, 14482 Potsdam, Germany\
}
\begin{document}

\maketitle

\begin{abstract}

The increasing precision in the determination of the Hubble parameter has reached a per cent level at which large-scale cosmic flows induced by inhomogeneities of the matter distribution become non-negligible. Here we use large-scale cosmological N-body simulations to study statistical properties of the local Hubble parameter as measured by local observers. We show that the distribution of the local Hubble parameter depends not only on the scale of inhomogeneities, but also on how one defines the positions of observers in the cosmic web and what reference frame is used. Observers located in random dark matter haloes measure on average lower expansion rates than those at random positions 
in space or in the centres of cosmic voids, and this effect is stronger from the halo rest frames compared to the CMB rest frame. 
We compare the predictions for the local Hubble parameter with observational constraints based on type Ia supernovae (SNIa) and CMB observations. Due to cosmic variance, for observers located in random haloes we show that the Hubble constant determined from nearby SNIa may differ from that measured from the CMB by $\pm0.8$ per cent at $1\sigma$ statistical significance. This scatter is too small to significantly alleviate a recently claimed discrepancy between current measurements assuming a flat $\Lambda$CDM model. However, for observers located in the centres of the largest voids permitted by the standard $\Lambda$CDM model, we find that Hubble constant measurements from SNIa would be biased high by 5 per cent, rendering this tension inexistent in this extreme case.

\end{abstract}


\section{Introduction}

The Hubble constant is arguably the most fundamental cosmological parameter. It is not only a measure of 
the expansion rate of the Universe, but it also sets the normalisation of the cosmic density parameters. Since 
the pioneering work by Edwin Hubble, all efforts made to determine this constant have been 
focused on improving the precision of distance measurements in the cosmic distance ladder 
as well as on reducing a number of systematic effects, related e.g. with the astrophysics of Cepheids and 
SNIa \citep[for a review see][]{Fre10}. These efforts have been enabled by the increasing amount and quality of the data, and resulted 
in measurements of the Hubble constant to an unprecedented per cent level of precision \citep{Rie11,Fre12,Tul13}.

The most competitive methods for the determination of the Hubble constant utilise combined measurements 
of distances to nearby Cepheids and SNIa, or CMB observations, although it is worth noting 
the growing relevance of constraints 
from time delays between gravitationally lensed multiple images of distant quasars \citep[see e.g.][]{Par10,Suy13}. 
The most recent measurement based on SNIa yields $H_{0}=73.8\pm2.4$ km s$^{-1}$ Mpc$^{-1}$ \citep[][]{Rie11}. 
On the other hand, assuming a flat $\Lambda$CDM cosmology a recent analysis of the CMB data from the \textit{Planck} mission 
leads to $H_{0}=67.88\pm0.77$ km s$^{-1}$ Mpc$^{-1}$ \citep{Ade13}, which differs from the previous result by $2.4\sigma$ 
\citep[although both measurements seem more compatible, when using the revised determination of $H_{0}$ 
from SNIa, based on an improved distance calibration; see][]{Efs13}. This raises the question to what degree this discrepancy is due to as yet unknown 
systematic uncertainties or large-scale inhomogeneities \citep{Mar13}. A similar problem concerns 
the difference in the expansion rate found between low- and high-redshift supernovae. The Hubble constant determined 
from SNIa within $75\hMpc$ was measured to be $6.5\pm1.8$ per cent higher than that at larger distances 
\citep{Jha07}. This effect, referred to as the Hubble bubble, is commonly ascribed to our location 
in an underdense region of the cosmic web \citep[][]{Zeh98}, although alternative solutions such as 
reddenning of local SNIa were also suggested \citep{Con07}. Most recent analyses of SNIa data reveal that a bulk 
flow of nearby SNIa prevails up to redshift $z\approx 0.06$, which corresponds to a comoving distance of $180\hMpc$ 
\citep[][]{Col11,Fei13}. It is worth noting that this scale of the bulk flow appears to coincide with the 
size of a plausible local underdensity determined from galaxy counts in the near-infrared \citep{Kee12,Whi13}.

The determination of the Hubble constant from measurements of distances and recessional velocities in 
the local Universe is inevitably affected by large-scale flows resulting from fluctuations in the matter 
distribution \citep{Cou13}. It is crucial then to theoretically predict these effects at the level of precision required 
by upcoming observations. Current theoretical works rely on analytical calculations based 
on either linear perturbation theory or its modifications that include some non-linear 
effects \citep[][]{Wan98,Coo06,LiSch08,Wie12,Kal13,Mar13}. Although this approach can give valid initial insights, 
it is insufficient for taking into account all non-linear effects and therefore to provide 
accurate predictions for the next generation measurements. Non-linear effects not only 
modify peculiar velocities at small scales, but they also implicitly define locations of reference 
observers in the most evolved structures of the cosmic web, i.e. dark matter (DM) haloes. The only way 
to include all these effects in a theoretical calculation of the variance of the perturbed Hubble flow 
is to use cosmological N-body simulations \citep[][]{Tur92,Mar09,Ara11}. 
In this paper, we utilise large-scale cosmological N-body simulations to study the effects of 
large-scale inhomogeneities and the distribution of DM haloes on the local measurements of the Hubble constant. 
The simulations were run in volumes comparable to the Hubble volume and thus 
they are suitable for a statistical study on the local Hubble flow. The goal of the paper is not only to quantify 
deviations of the local Hubble parameter from the global expansion rate, 
but also to consider a number of assumptions, such as the location of reference observers in the cosmic web 
or the reference frame for the measurement, which considerably modify these predictions.

The manuscript is organised as follows. In section 2, we describe the simulations and methods used 
to find DM haloes and cosmic voids. In the next section, we introduce the notion of the local Hubble 
parameter and its connection to large-scale inhomogeneities. We also describe here the technical 
details of calculating predictions for the local Hubble parameter based on cosmological simulations. 
Section 4 presents the main results of the paper and is divided into several detailed 
subsections in which the effects of different sets of assumptions to calculate the local Hubble 
parameter are discussed. In section 5, we calculate the theoretical predictions for a few different measurements of the 
Hubble parameter based on SNIa or CMB data and compare them to observational constraints. We conclude 
and summarise in section 6.

\section{Simulations}

We use two large-scale cosmological N-body simulations, 
the JUropa HuBbLE volumE\footnote{http://jubilee-project.org} \citep[Jubilee,][]{Wat13} with a volume of $(6\hGpc)^{3}$ and 
a part of the Big MultiDark (Big MD, He{\ss} et al., in prep.) 
which is a suite of simulations with volumes of $(2.5\hGpc)^{3}$. 
The two simulations are based on the results of the of 5-year CMB data 
from the \textit{Wilkinson Microwave Anisotropy Probe} (WMAP) satellite \citep[][]{Dun09,Kom09}. 
The assumed cosmological parameters, $(\Omega_{m}=0.27,
\Omega_{\Lambda}=0.73,\Omega_{b}=0.044,h=0.70,\sigma_{8}=0.80,n_{s}=0.96$) for the Jubilee and 
$(\Omega_{m}=0.29,\Omega_{\Lambda}=0.71,\Omega_{b}=0.047,h=0.70,\sigma_{8}=0.82,n_{s}=0.95$) 
for the MultiDark, are consistent with recent results from \textit{Planck} \citep[][]{Ade13}. For more details 
of each simulation see \citet{Wat13} and He{\ss} et al. (in preparation).

The two simulations have $6000^{3}$ (Jubilee) and $3840^{3}$ (Big MD) particles. The corresponding 
particles masses are $7.49\times10^{10}\hMsun$ and $2.2\times10^{10}\hMsun$, yielding minimum 
resolved halo masses (with $50$ particles) of $4\times10^{12}\hMsun$ and $1\times10^{12}\hMsun$, respectively. 
Combining the two simulations, we resolve DM haloes spanning the mass range from Milky-Way-size 
galaxies  with $10^{12}\hMsun$ to massive galaxy clusters with $10^{15}\hMsun$. 

DM haloes in both simulations are found using the {\sevensize\bf AMIGA} halo finder, {\sevensize\bf AHF} 
\citep[][]{Gil04,Kno09}. The finder identifies the halo centres as the overdensity peaks located on a recursively 
refined grid.  Every centre is then used to find gravitationally bound particles and, on the basis of these particles, 
to calculate various properties of the haloes, including the halo mass $M_{\rm halo}$ which is defined as the mass of a spherical 
overdensity region with the mean density equal to $178\rho_{b}$, where $\rho_{b}$ is the background density. 
The final sets of all distinct haloes identified in the Jubilee and the Big MD simulations (both at redshift $z=0$) 
contain $9.1\times10^{7}$ haloes with masses $M_{\rm halo}>10^{13}\hMsun$ and $5.8\times10^{7}$ haloes with masses 
$M_{\rm halo}>10^{12}\hMsun$, respectively. For the latter, there are $5\times10^{7}$ haloes with masses 
$10^{12}\hMsun<M_{\rm halo}<10^{13}\hMsun$.

In addition to DM haloes, we also identify voids formed in the Jubilee simulation. Voids are found as regions 
in the simulation box which do not contain DM haloes above a certain mass \citep{Wat13}. We adopt $10^{14}\hMsun$ 
as the mass threshold of the void finder. With this mass limit, we identify the most extended voids which have the strongest 
effect on the local Hubble parameter. Every void is defined as a sphere 
that maximally fills a volume devoid of haloes \citep[][]{Got03}. The centre of this sphere is used as the void centre. 
The total number of voids found at redshift $z=0$ in the $6\hGpc$ simulation is $2.6\times10^{5}$.

\section{Local Hubble parameter from simulations}

The observed velocities of galaxies in the local Universe combine a recessional velocity component due to the global expansion 
of the Universe and a peculiar velocity component resulting from the local density fluctuations. Without 
prior information on the distances to galaxies, these two velocity components cannot be disentangled, and therefore 
local measurements of the Hubble constant may differ from the actual global rate of expansion. 
The local Hubble parameter depends not only on the cosmological model, but also on the location of the 
observer in the cosmic web and on the selection of the objects used for the measurement. A deviation from the global 
expansion rate is expected on distances which are comparable in size or smaller than cosmic voids, which set 
a natural scale of transition to homogeneity \citep{Scr12}. To study the effects of inhomogeneities on the local 
Hubble parameter, cosmological simulations should be run in boxes much larger than the largest 
cosmic structures, i.e. cosmic voids. Depending on the threshold for the density contrast, 
the effective radii of voids (the radii of spheres with enclosed volumes equal to the volumes of voids) span the range 
from $10\hMpc$ to $100\hMpc$ \citep{Wat13}. Due to asphericity, maximum sizes of these objects are even 
a few times larger than the effective radii. Therefore, the side length of the simulation box should 
be at least of a few $\hGpc$.

The local Hubble parameter $H_{{\rm loc}}$ is the slope of a linear relation between the observed velocities of 
galaxies or groups of galaxies and the distances from the observer to these objects
\begin{equation}
\mathbf{v}_{\rm pec}\cdot\mathbf{\hat{r}}+H_{0}r=H_{{\rm loc}}r,
\label{local_Hubble_flow}
\end{equation}
where $\mathbf{v}_{\rm pec}$ is the peculiar velocity vector, $\mathbf{\hat{r}}$ is the normalised position 
vector and $r$ is the distance. In a homogeneous universe, peculiar velocities vanish and 
one recovers the Hubble law, i.e. $H_{{\rm loc}}=H_{0}$. In a non-homogeneous universe, observers located in 
voids or overdense regions measure expansion rates that are larger or smaller than $H_{0}$, respectively. 

The calculation of the local Hubble flow from theory can be expressed in Bayesian terms. For a fixed cosmological 
model, the probability distribution of the local Hubble parameter is given by
\begin{equation}
p(H_{{\rm loc}})=p(H_{{\rm loc}}|\textrm{observer})p(\textrm{observer}),
\end{equation}
where $p(\textrm{observer})$ describes the probability distribution of the locations of observers in the cosmic web. 
For simplicity, this formula neglects a number of factors related to how $H_{\rm loc}$ is actually measured in different 
kinds of observations. Among these factors, the most important are the selection function of objects used for the measurement 
(both along the line of sight and on the sky) and the errors on distances. In section 4, we assume an idealised measurement for which all errors are negligible and the 
observations are complete at all radii $r<r_{\rm max}$. This means that the probability $p(H_{\rm loc})$ is fully determined by the 
large-scale structures formed in the simulations and the location of the observers in the cosmic web. The effect of 
adopting a selection function for the redshift distribution is considered 
in section 5, where we calculate probability distributions for the differences in $H_{\rm loc}$ measured at 
different scales.

The probability $p(\textrm{observer})$ should encompass a priori information about our own location 
in the cosmic web, such as that we live in a galaxy group within a DM halo of a certain mass. This kind of 
information can significantly modify the final predictions for $H_{{\rm loc}}$. 
The probability distribution $p(\textrm{observer})$ may also be interpreted as a mathematical representation of the 
Copernican principle or any deviation from it. It is a way of quantifying the meaning of the term `typical observers'. 
From a technical point of view, we shall not deal with any explicit form of $p(\textrm{observer})$, but instead we shall consider 
different schemes for placing random observers in the cosmic web. Every scheme, however, corresponds to a unique form of 
$p(\textrm{observer})$.

In order to calculate $p(H_{\rm loc})$ from the simulations, we define a set of observers and haloes hosting observable 
objects in the Universe, i.e. galaxies and groups of galaxies. In all cases, we use $10^{5}$ observers which are randomly 
distributed in the cosmic web according to different choices for $p(\textrm{observer})$. We checked that 
the number of observers is sufficient to precisely calculate $p(H_{\rm loc})$. For every observer, we find all haloes 
at distances $r<r_{\rm max}$ and fit the linear model of eq.~(\ref{local_Hubble_flow}) to the line-of-sight velocities and distances. 
The resulting $10^{5}$ best-fit values of $H_{{\rm loc}}$ are then treated as a random sample 
drawn from the probability distribution $p(H_{\rm loc})$. We use these values to compute the mean and the confidence intervals 
of the probability distribution $p(H_{\rm loc})$. We repeat the calculations for the maximum radius $r_{\rm max}$ spanning the range from 
$20\hMpc$ to $300\hMpc$. The linear fit is carried out assuming equal weights for all the haloes within $r_{\rm max}$ so that the 
final probability distribution of the local Hubble parameter is fully determined by the large scale structures that emerged from the 
evolution of the assumed cosmological model. Unless it is explicitly stated (as in subsection 4.5), the peculiar velocities in 
eq.~(\ref{local_Hubble_flow}) are given by the bulk velocities of the haloes in the comoving rest frame. This choice of 
the reference frame corresponds to measuring the Hubble parameter in the CMB rest frame \citep{Tur92}. In observations 
of nearby Cepheids or SNIa, the transformation to this reference frame relies on correcting the observed redshifts for the 
motion of the Milky Way as determined from the CMB dipole.

When we fit for the local Hubble parameter, we use the peculiar velocities of the haloes in the $z=0$ snapshots of the simulations. 
However, the assumed maximum value of $r_{\rm max}$, $300\hMpc$, corresponds to redshift $z=0.1$. This means that our calculation 
neglects the evolution of peculiar velocities at $z<0.1$. Using linear perturbation theory (see more details in subsection 4.1), 
we estimate the impact of the evolution between $z=0$ and $z=0.1$, and find that the resulting relative corrections to the variance of 
the local Hubble parameter are smaller than $5$ per cent. We also check that the corrections needed to account for the 
differences in cosmological parameters of the two simulations are of the same order of magnitude. To prevent our results from 
being affected by these two effects, we provide all our results within a $10$ per cent precision.

\section{Results}
Here we consider several choices for $p({\rm observer})$ and calculate the corresponding distribution of $H_{\rm loc}$. 
We explore the following distributions of observers within the simulation box: random in space; 
random in DM haloes of different masses; in the centres of voids; and in the local rest frames of DM haloes.

\subsection{Random observers in space}

The first simple way of selecting observers is to draw random positions in the simulation box \citep{Tur92}. In this case, 
observers are not assigned to the haloes and the resulting distribution of the local Hubble parameter represents 
a volume weighted statistic, i.e. one assigns equal probabilities to all positions in the simulation box rather than 
to all distinct haloes. This scheme of distributing observers is less natural than selecting random DM haloes as 
the locations for observers. However, the main motivation for considering this approach is that the 
same distribution of observers is implied when using analytical calculations based on perturbation theory (for which there 
is no prediction for halo formation and thus all statistical quantities are weighted by volume rather than by haloes). 
Therefore, the framework considered here allows us to directly compare the results from the simulations to those from analytical 
calculations \citep[see e.g.][]{Wan98,Mar13}. For fitting the Hubble relation given by eq.~(\ref{local_Hubble_flow}), 
we use haloes with masses $M_{\rm halo}>10^{13}\hMpc$ from the $6\hGpc$ simulation box.

\begin{figure}
\begin{center}
    \leavevmode
    \epsfxsize=8cm
    \epsfbox[65 65 560 408]{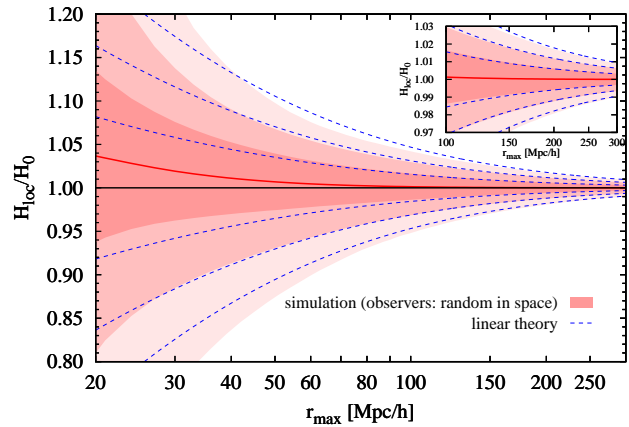}
\end{center}
\caption{Probability distribution of the local Hubble parameter $H_{{\rm loc}}$ within the radius $r_{\rm max}$, 
as measured by observers randomly distributed in space of the $6\hGpc$ simulation box (red shaded contours). 
The red solid line is the mean and the contours are the $68.3$, $95.4$ and $99.7$ per cent confidence intervals. 
For comparison purposes, the blue dashed contours show the corresponding confidence intervals calculated 
using linear perturbation theory, with the mean value equal to 1. The inset panel zooms into the 
main plot for radii $r_{\rm max}>100\hMpc$.}
\label{linear-comp}
\end{figure}

Fig.~\ref{linear-comp} shows the confidence intervals of the probability distribution of 
$H_{\rm loc}$ with respect to its global value $H_{0}$ as a function of the 
maximum radius of observations $r_{\rm max}$ (red shaded contours). As expected, the measurement of the Hubble parameter converges to $H_{0}$ at large distances. The 
relative deviation from the global Hubble flow is as small as $1$ per cent within $150\hMpc$. Note also 
that for $r_{\rm max}<50\hMpc$ the mean value of $H_{\rm loc}/H_{0}$ is larger than 1 by a few per cent. 
This effect results from using a volume-weighted statistic which 
enhances the contribution of the cosmic outflows inside volume-dominated voids.

It is interesting to compare our results from cosmological simulations with analytical calculations based on linear 
perturbation theory. According to this theory, the variance of the local Hubble parameter measured within a sphere 
with radius $r_{\rm max}$ is given by the following equation \citep[see e.g.][]{Wan98}
\begin{equation}
\Big\langle\Big(\frac{H_{{\rm loc}}-H_{0}}{H_{0}}\Big)^{2}\Big\rangle=\frac{(\Omega_{m}^{0.55})^{2}}{2\pi^{2}r_{\rm max}}\int_{0}^{\infty}P(k)[f(x)/x^{2}]\textrm{d}k,
\label{linear-var}
\end{equation}
where $\Omega_{m}$ is the matter density parameter at the present time, $x=kr_{\rm max}$, $P(k)$ is the power spectrum of the matter density fluctuations and 
\begin{equation}
f(x)=\frac{3}{x^{2}}\Big(\sin x-\int_{0}^{x}\frac{\sin y}{y}\textrm{d}y\Big).
\end{equation}
For $\Lambda$CDM, $\Omega_{m}^{0.55}$ is an approximation for the linear growth rate of density perturbations $\delta(a)$, 
i.e. $\textrm{d}\ln\delta/\textrm{d}\ln a\simeq\Omega_{m}(a)^{0.55}$ \citep{Lin05}, where $a$ is the cosmic scale factor.

The blue dashed contours in Fig.~\ref{linear-comp} show the confidence intervals for a Gaussian probability distribution with 
the standard deviation given by eq.~(\ref{linear-var}) for the same cosmological parameters and 
power spectrum as used in the simulation. The linear approximation recovers the results from the simulations at scales 
$r_{\rm max}\gtrsim 100\hMpc$. Non-linear evolution effects, such as the mean value being larger than 1, become 
relevant at scales corresponding to typical sizes of voids, i.e. $r_{\rm max}\lesssim 40\hMpc$. As shown by \citet{Mar13}, 
the linear approximation can be analytically corrected for the non-linear evolution of cosmic voids by assuming a log-normal 
probability distribution instead of a Gaussian.

For the various cases presented in this section, Table~\ref{intervals} lists the mean and scatter of the relative difference between the local $H_{\rm loc}$ and the global 
$H_{0}$ Hubble parameter within three maximum radii $r_{\rm max}$: $50\hMpc$, $75\hMpc$ and $150\hMpc$. 
The maximum radius $r_{\rm max}=75\hMpc$, corresponding to the Hubble velocity 
$cz_{\rm max}=7500$ km~s$^{-1}$, is commonly adopted as the limiting value separating low-$z$ 
SNIa, whose Hubble diagram is likely to be significantly perturbed by local inhomogeneities 
\citep{Zeh98,Jha07}, from high-$z$ ones, which are thought to probe the global expansion of the Universe 
\citep{Rie11}.

\begin{table*}
\begin{center}
\begin{tabular}{cc r @ {.} l c  r @ {.} l c r @{.} l c}
observers & $\log_{10}M_{\rm halo}[\hMsun]$ & 
\multicolumn{2}{c}{$\mu_{50}[\%]$} & $\sigma_{50}[\%]$ 
& \multicolumn{2}{c}{$\mu_{75}[\%]$} & $\sigma_{75}[\%]$ 
& \multicolumn{2}{c}{$\mu_{150}[\%]$} & $\sigma_{150}[\%]$ \\
\hline
random in space                           & $>13$ & 0 & 7 & 3.3 & 0 & 3 & 2.1 & 0 & 0 & 0.9 \\
random in haloes                          & $ >13$ & -1 & 7 & 4.0 & -0 & 8 & 2.4 & -0 & 2 & 0.9 \\
centres of voids                             &  $>13$ & 2 & 5 & 2.5 &1 & 0 & 1.8 & 0 & 1 & 0.9 \\
random in haloes                          & $(12,13)$ & -0 & 6 & 3.3 & -0 & 3 & 2.1 & 0 & 0 & 0.9 \\
random in space                           & no haloes/linear model & 0 & 0 & 3.5 & 0 & 0 & 2.2 & 0 & 0 & 0.9 \\
random in haloes/halo rest frame &  $>13$ & -3 & 9 & 4.7 & -1 & 9 & 2.7 & -0 & 4 & 1.0 \\
\end{tabular}
\caption{Mean $\mu$ and scatter $\sigma$ of the relative difference between the local and global 
expansion rates, i.e. $\mu=\langle H_{\rm loc}/H_{0}-1\rangle$ and $\sigma=\langle (H_{\rm loc}/H_{0}-1-\mu)^{2}\rangle^{1/2}$. 
The first and second columns describe the selection of observers and the mass range of the haloes used in the calculation. 
The remaining columns list the mean and the scatter of $(H_{\rm loc}-H_{0})/H_{0}$ for $r_{\rm max}$ of $50\hMpc$, $75\hMpc$ and $150\hMpc$, respectively.}
\label{intervals}
\end{center}
\end{table*}

\subsection{Observers in random haloes}

The most natural way of distributing observers in the cosmic web is by positioning them in random DM haloes. 
This scheme assumes that typical observers are located in structures which are 
embedded in DM haloes, i.e. galaxies or groups of galaxies, such as our own location in the Local Group. 
This substantially decreases the effective volume of space populated by observers. In this scheme, observers 
tend to occupy overdense regions and thus their measurement of the Hubble constant is mostly affected by large scale infall.

\begin{figure}
\begin{center}
    \leavevmode
    \epsfxsize=8cm
    \epsfbox[65 65 560 408]{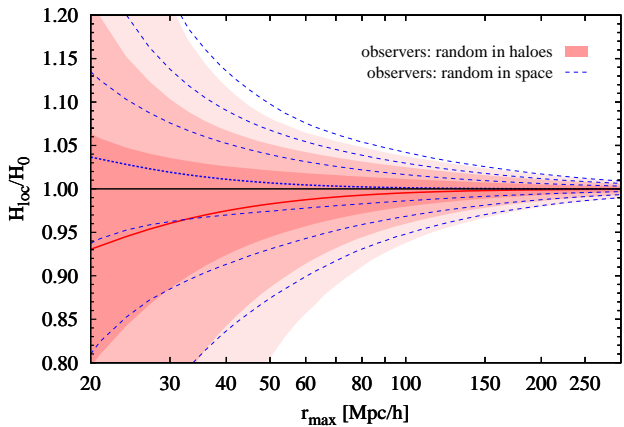}
\end{center}
\caption{Probability distribution of the local Hubble parameter $H_{{\rm loc}}$ within the radius $r_{\rm max}$, 
as measured by observers randomly distributed in DM haloes from the $6\hGpc$ simulation box (red shaded contours), 
and compared to those for the case with observers randomly distributed in space (blue dashed contours). 
For each case, the contours show the $68.3$, $95.4$ and $99.7$ per cent confidence intervals, whereas the 
red solid and blue dotted lines show the mean values.
}
\label{halo}
\end{figure}

\begin{figure}
\begin{center}
    \leavevmode
    \epsfxsize=8cm
    \epsfbox[65 65 560 408]{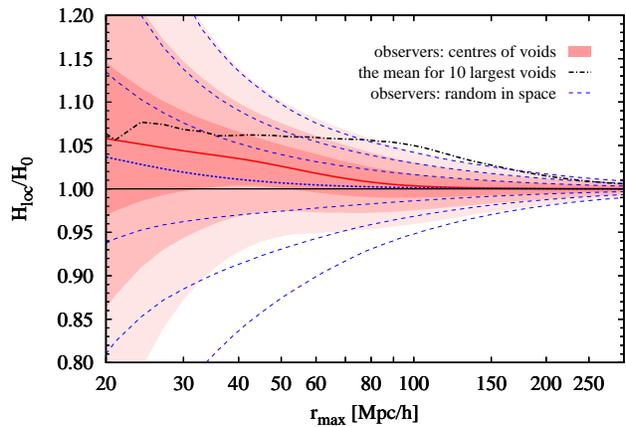}
\end{center}
\caption{Probability distribution of the local Hubble parameter $H_{{\rm loc}}$ within the radius $r_{\rm max}$, 
as measured by observers located in the centres of voids found in the $6\hGpc$ simulation box (red shaded contours), 
and compared to those for the case with observers randomly distributed in space (blue dashed contours). 
For each case, the contours show the $68.3$, $95.4$ and $99.7$ per cent confidence intervals, whereas the 
red solid and blue dotted lines show the mean values. The black dash-dotted line shows instead the mean values 
obtained when only the centres of the 10 largest voids (in terms of the size) are used.}
\label{void}
\end{figure}

Fig.~\ref{halo} shows the confidence intervals for the local Hubble parameter as measured by observers located 
in randomly selected DM haloes with masses $M_{\rm halo}>10^{13}\hMsun$ in the $6\hGpc$ simulation (red shaded contours). 
The same minimum mass of $10^{13}\hMsun$ is assumed for haloes used in fitting the linear Hubble relation. These results 
are compared to the previous case for observers randomly distributed space (blue dashed 
contours corresponding to the red shaded contours in Fig.~\ref{linear-comp}).

For scales $r_{\rm max}\lesssim 70\hMpc$, the distribution of deviations of $H_{\rm loc}$ from $H_{0}$ differs 
significantly from that of the case with observers randomly distributed in space. At these scales, the probability distribution of 
$H_{\rm loc}$ is skewed towards smaller values and the mean local Hubble parameter is less than $H_{0}$. 
This is due to the fact that observers populate preferentially overdense, infall-dominated regions. The size of the scatter at 
$r_{\rm max}=75\hMpc$ is comparable to that for observers randomly distributed in space, but 
the confidence intervals are shifted towards smaller values by $20$ per cent of their size. 
When compared to Fig.~\ref{linear-comp}, it appears that the selection of observers by haloes has \textit{a significantly larger} effect 
on the probability distribution of $H_{\rm loc}$ than the corrections due to non-linear evolution effects for observers 
randomly distributed in space.

\subsection{Observers in the centres of voids}

The motivation for considering the centres of voids as the locations for the observers comes from observations 
of SNIa. The Hubble constant appears to be slightly larger at small compared to large distances \citep{Zeh98,Jha07}. 
This trend of the Hubble parameter with the distance, referred to as the Hubble bubble, is statistically significant 
and is commonly ascribed to a local void. The location of the Local Group in a void leads to another choice of the prior probability 
$p(\textrm{observer})$: observers located in voids. Here we consider a rather extreme possibility and locate the observers in the 
centres of voids.

We use voids from the $6\hGpc$ simulation that are identified as regions devoid of haloes with masses $M_{\rm halo}>10^{14}\hMsun$. The total 
number of voids is of the same order of magnitude as the adopted number of observers, i.e. $10^{5}$. 
Fig.~\ref{void} shows the confidence intervals of the resulting probability distribution of $H_{\rm loc}$ 
as a function of $r_{\rm max}$ (red shaded contours), compared to 
the results for observers randomly distributed in space (blue dashed contours). From this figure one can see that 
for observers located in voids the local Hubble parameter is affected 
by large scale outflow. For this case, the mean value of $H_{\rm loc}$ decreases from $1.06H_{0}$ at 
$r_{\rm max}=20\hMpc$ to $1.01H_{0}$ at $r_{\rm max}=75\hMpc$. Note also that 
the confidence intervals here are more similar to those for the case with observers randomly distributed in space 
rather than in DM haloes. This shows how strong the contribution from voids is for a volume-weighted statistic of the 
local Hubble parameter.

As the most extreme example, we also plot the mean local Hubble parameter as measured with respect to the centres of the 10 largest voids 
selected by size, with radii between $90\hMpc$ and $100\hMpc$ (black dash-dotted line). In this case, the mean of $H_{\rm loc}$ is $\sim5$ per cent larger than $H_{0}$ within all radii up to $\sim100\hMpc$. 
Then, it drops to $\sim1$ per cent of $H_{0}$ within $\sim200\hMpc$. At scales between $90\hMpc$ and  $150\hMpc$, the local Hubble 
parameter reaches the upper limit of the $99.7$ per cent confidence interval of the probability distribution obtained for observers distributed in the 
centres of all voids.

\subsection{Observers in groups or clusters of galaxies}

Massive haloes tend to populate denser environments, which are surrounded by large scale infall. Therefore, when selecting 
observers by haloes one should consider the dependence on the halo mass. We show this effect by considering haloes with masses 
$10^{12}\hMsun<M_{\rm halo}<10^{13}\hMsun$ from the $2.5\hGpc$ simulation box. These haloes correspond to massive galaxies or galaxy groups, 
with the mass range including the mass of the Local Group estimated at $5\times10^{12}\hMsun$ \citep[][]{Li08,Par13}. 
This halo population may be regarded as a random sample 
drawn from the most natural prior probability of $p(\textrm{observer})$ describing observers resembling our location in the local cosmic 
web (we belong to a group, but not to a cluster). We use the same halo population to both select observers and fit the linear Hubble relation.

\begin{figure}
\begin{center}
    \leavevmode
    \epsfxsize=8cm
    \epsfbox[65 65 560 408]{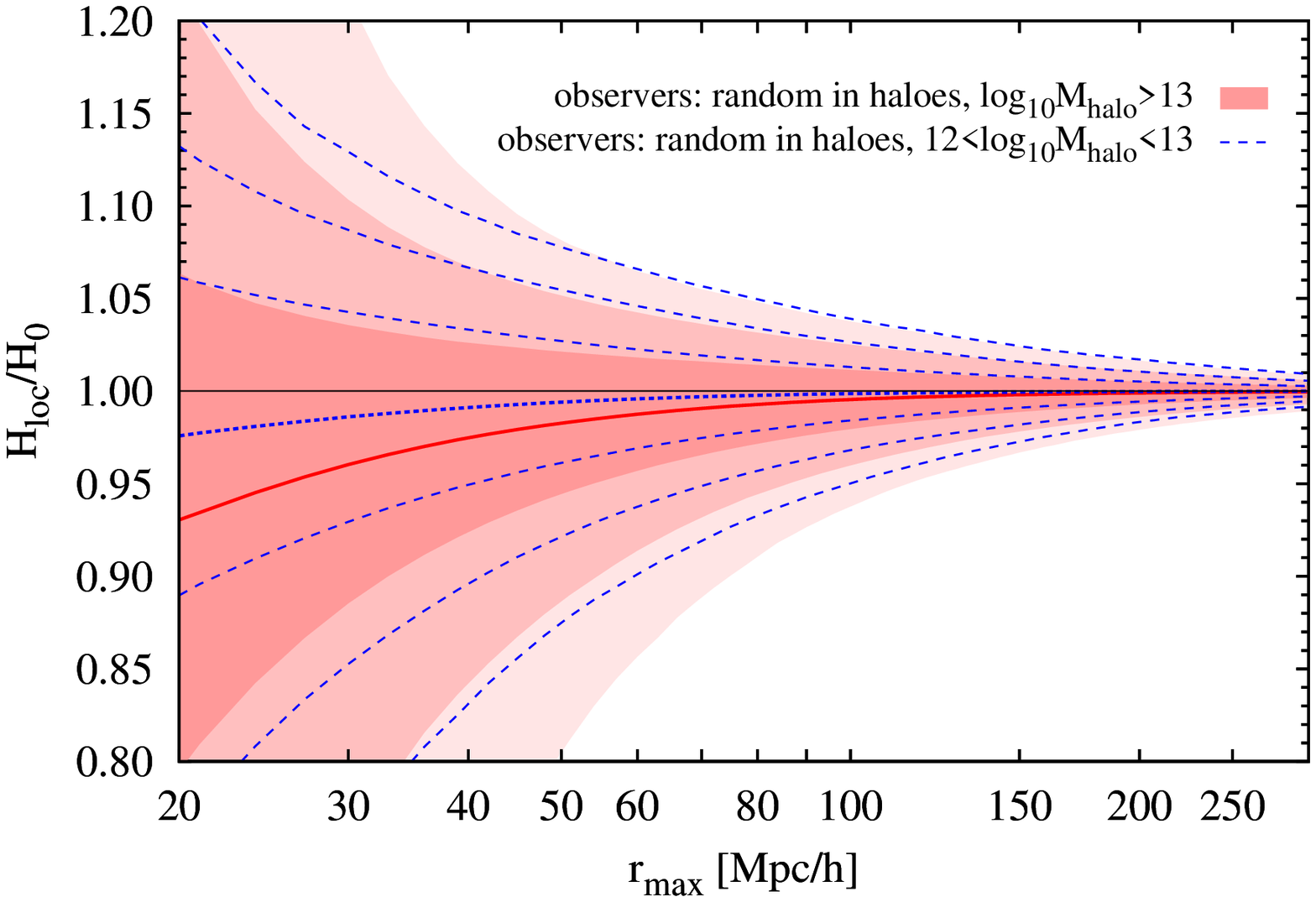}
\end{center}
\caption{Probability distribution of the local Hubble parameter $H_{{\rm loc}}$ within the radius $r_{\rm max}$, 
as measured by observers randomly distributed in DM haloes for two different halo populations: haloes with masses 
$M_{\rm halo}>10^{13}\hMsun$ from the $6\hGpc$ simulation box (red shaded contours) and with $10^{12}\hMsun<M_{\rm halo}<10^{13}\hMsun$ 
from the $2.5\hGpc$ simulation box (blue dashed contours). For each case, the contours show the $68.3$, $95.4$ and $99.7$ per cent confidence intervals, 
whereas the red solid and blue dotted lines show the mean values.}
\label{halo-halo}
\end{figure}

Fig.~\ref{halo-halo} shows the probability distributions of $H_{\rm loc}$ as measured by observers located in 
randomly selected Local-Group-like haloes, i.e. haloes with masses $10^{12}\hMsun<M_{\rm halo}<10^{13}\hMsun$ 
(blue dashed contours). This distribution is compared to the case for observers 
located in massive haloes of $M_{\rm halo}>10^{13}\hMsun$ from the $6\hGpc$ simulation box (red shaded contours, the same as in Fig.~\ref{halo}). 
It can be seen from this figure that the local Hubble flow is less affected by the large-scale infall 
when using less massive haloes. As expected, the differences due to halo masses become negligible at large scales. 
Comparing to Fig.~\ref{linear-comp}, one can see that the confidence intervals of $H_{\rm loc}/H_{0}$ at $r_{\rm max}<75\hMpc$ for observers 
located in less massive haloes are more similar to those obtained with simple analytical calculations based on linear perturbation theory 
than those for observers in more massive haloes (see also Table~\ref{intervals}).

\subsection{Reference frame}

Until now, we have assumed that observers are at rest in the comoving coordinate system. 
This means that the velocities used for fitting the Hubble relation of eq.~(\ref{local_Hubble_flow}) are measured in the CMB rest frame, 
as in the standard practice of transforming the observed redshifts to the CMB rest frame 
by using the bulk velocity of the Milky Way as determined from the CMB dipole. Here we consider 
a hypothetical situation in which velocities are measured instead in the local rest frames of the haloes.

\begin{figure}
\begin{center}
    \leavevmode
    \epsfxsize=8cm
    \epsfbox[65 65 560 408]{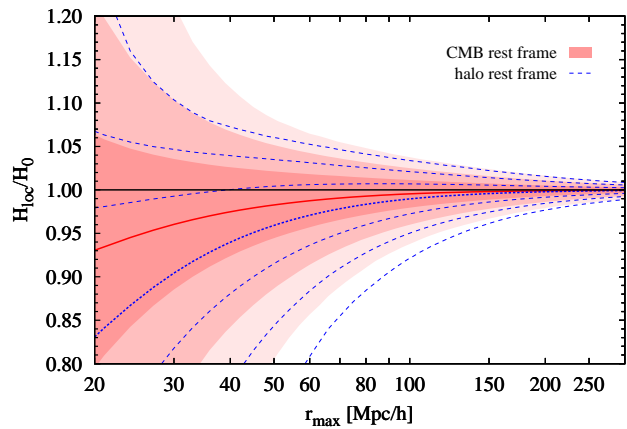}
\end{center}
\caption{Probability distribution of the local Hubble parameter $H_{{\rm loc}}$ within the radius $r_{\rm max}$, 
as measured by observers in the CMB rest frame (red shaded contours) or the local rest frames of DM haloes (blue dashed contours). 
In both cases, observers are located in random haloes with masses $M_{\rm halo}>10^{13}\Msun$ selected from the $6\hGpc$ simulation box. 
The contours show the $68.3$, $95.4$ and $99.7$ per cent confidence intervals, whereas the red solid and blue dotted lines show the mean values.}
\label{rest-frame}
\end{figure}

Fig.~\ref{rest-frame} compares the predictions for the measurement of the local Hubble parameter in two reference frames: 
the CMB rest frame (red shaded contours, the same as in Fig.~\ref{halo}) 
and the local halo rest frames (blue dashed contours). In both cases, observers are located in random haloes with masses 
$M_{\rm halo}>10^{13}\hMsun$ from the $6\hGpc$ simulation. The figure shows that the change of reference frame has a significant 
effect. At scales of $r_{\rm max}\lesssim 80\hMpc$, the mean local Hubble parameter measured in the halo rest frames  
is smaller than the mean determined in the CMB rest frame by $50-70$ per cent of the intrinsic scatter. 
Among all effects summarised in Table~\ref{intervals}, the change of the reference frame is the largest.

\section{Comparison with observations}

To compare the probability distribution of the local Hubble parameter to realistic observations, 
one needs to account for the incompleteness in the observational selection of the objects used to trace the Hubble flow, as e.g. SNIa. 
A convenient way to include the selection in redshift space is to assign distance-dependent weights $w(r)$ to the haloes used for 
fitting the Hubble flow. The local Hubble constant is then given by a weighted estimator of the following form
\begin{equation}
H_{{\rm loc}}=\sum_{i}w(r_{i})(\mathbf{v_{{\rm pec}\;i}}\mathbf{\hat{r}_{i}}+H_{0}r_{i})r_{i}/\sum_{i}w(r_{i})r_{i}^{2}.
\end{equation}
The weights in this equation are the ratios of the density of supernovae to the density of DM haloes at comoving 
distances $r_{i}$. The density of the observed supernovae can be easily obtained by converting their redshift distribution 
to the density in the comoving coordinates. For this conversion, we assume $H_{0}=70$ km~s$^{-1}$~Mpc$^{-1}$.

\begin{figure}
\begin{center}
    \leavevmode
    \epsfxsize=8cm
    \epsfbox[65 65 560 408]{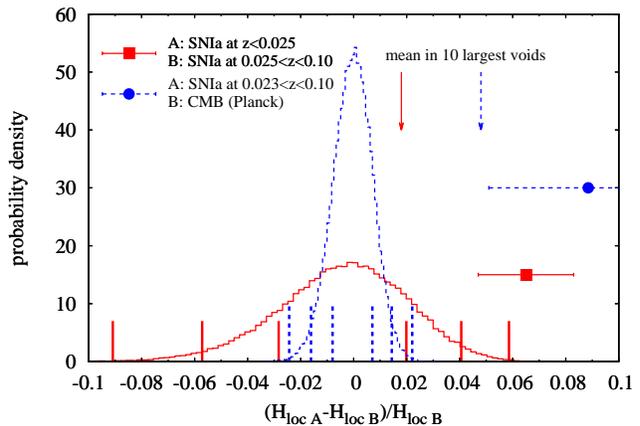}
\end{center}
\caption{Relative differences in $H_{\rm loc}$ measured using two sets of observations, A and B, 
that probe the expansion rate at different scales. The blue point with dashed error bars shows the difference between 
the expansion rates inferred from SNIa at redshifts $z<0.025$ (A) and $0.025<z<0.10$ (B) from \citet{Jha07}. 
The red point with solid error bars shows the relative difference between $H_{\rm loc}$ determined from nearby SNIa at redshifts 
$0.023<z<0.10$ \citep[A;][]{Rie11} and CMB measurements from \textit{Planck} \citep[B;][]{Ade13}. 
The error bars represent only the statistical uncertainties of the measurements. The corresponding lines 
show the probability distributions of the predicted relative differences in the expansion rate, 
as measured by observers located in DM haloes with masses $10^{12}\hMsun<M_{\rm halo}<10^{13}\hMsun$ 
formed in a cosmological simulation of a standard $\Lambda$CDM model (the $2.5\hGpc$ box). 
The vertical lines indicate the $68.3$, $95.4$ and $99.7$ per cent confidence intervals. 
The arrows show the mean relative differences in $H_{\rm loc}$ for observers 
located in the centres of the 10 largest voids from the $6\hGpc$ simulation.
}
\label{pdf-diff}
\end{figure}

We adopt the redshift distributions of three different samples of nearby SNIa used in measurements 
of the Hubble parameter. The first and second samples were compiled by \citet{Jha07} and have 
supernovae at redshifts $z<0.025$ and $0.025<z<0.10$, respectively. These two data sets revealed that the expansion 
rate of closer supernovae is higher than the rate determined from the distant ones \citep[see also][]{Zeh98}. 
The third sample includes all supernovae from \citet{Hic09} at redshifts $0.023<z<0.10$. This is nearly 
the same data compilation that was used in the measurement of $H_{0}$ by \citet{Rie11}.

We calculate the probability distributions of $H_{{\rm loc}}$ using the redshift distributions of  these three SNIa samples. 
The reference observers are randomly distributed in DM haloes with masses 
$10^{12}\hMsun<M_{\rm halo}<10^{13}\hMsun$ from the $2.5\hGpc$ simulation box and the local Hubble flow is measured in 
the rest frame of the CMB. Comparison with observational constraints on $H_{{\rm loc}}$ can only be made 
in terms of the relative differences between the Hubble parameters determined 
from data sets probing $H_{\rm loc}$ at different scales. 
Here we consider two combinations of measurements which have recently attracted considerable attention: 
i. the difference in the determination of $H_{{\rm loc}}$ between the $z<0.025$ and $0.025<z<0.10$ SNIa samples 
reported by \citet{Jha07} and ii. the difference between the expansion rate 
determined by using nearby SNIa with an improved 
distance calibration from Cepheids \citep{Rie11} and that by using CMB observations 
from \textit{Planck} \citep{Ade13}. In the latter case, we assume that the Hubble constant measured from the CMB is 
not affected by large-scale structures, i.e. $H_{{\rm loc}}=H_{0}$.

Fig.~\ref{pdf-diff} shows the predicted probability distributions of the relative differences in $H_{{\rm loc}}$ for 
the two combinations of observations (red solid and blue dashed lines, respectively) as well as the corresponding 
measurements from \citet{Jha07}, $6.5\pm1.8$ per cent, and from combining results from SNIa \citep{Rie11} and 
the CMB \citep{Ade13}, $8.9\pm3.7$ per cent. All quoted errors include only the statistical uncertainties of the measurements.

The $3.6\sigma$ tension between the expansion rates obtained from nearby ($z<0.025$) and 
distant ($0.025<z<0.10$) SNIa can be easily alleviated by assuming the existence of a local void. For this case, the 
$99.7$ per cent interval of the predicted relative difference in $H_{{\rm loc}}$ reaches $5.8$ per cent. 
Taking into account the dispersion of the theoretical distribution in the error budget yields a $2.2\sigma$ 
deviation of the observational results from the predictions of the $\Lambda$CDM model assumed here, i.e. $6.9\pm3.0$ per cent 
(with the mean of the theoretical probability distribution equal to $-0.4$ per cent).

The effect of cosmic variance is too small to reduce the tension between the value of the Hubble constant 
determined from nearby SNIa and that from the CMB. The standard deviation of the theoretical probability distribution 
is $0.8$ per cent, which is $30$ per cent smaller than the value obtained in analytical calculations by \citet{Mar13}. 
This value has a negligible effect on the error budget of the relative difference in the expansion rates obtained from SNIa 
and the CMB, for which the tension remains unchanged at $2.4\sigma$ statistical significance (neglecting all systematic errors). 
We note, however, that this conclusion strongly relies on the assumed prior probability $p(\textrm{observer})$. 
 Although we chose a very conservative form for $p(\textrm{observer})$ (random DM haloes with masses comparable to the mass 
 of the Local Group), it is interesting to consider other possibilities. As an extreme example, we recalculate the local Hubble 
 parameter using observers placed in the centres of the 10 largest voids from the $6\hGpc$ simulation. The mean relative 
 difference in $H_{\rm loc}$ from SNIa and the CMB is $4.8$ per cent (see the blue arrow in Fig.~\ref{pdf-diff}), which accounts 
 for $55$ per cent of the measured value and thus decreases the tension between the two measurements to $1\sigma$. 
 On the other hand, the mean relative difference 
 between $H_{\rm loc}$ from $z<0.025$ and $0.025<z<0.10$ SNIa is $1.6$ per cent (see the red arrow in Fig.~\ref{pdf-diff}). 
 Such a small value results from the fact that the local Hubble parameter is nearly independent of radius within the range of 
 distances to nearby SNIa (see the black profile in Fig.~\ref{void}; the median of the distances is $75\hMpc$ and the 
 distribution strongly decays at larger radii).

\section{Summary and conclusions}

We made use of large-scale cosmological N-body simulations to study the effects of inhomogeneities and the distribution of 
DM haloes on the measurement of the local Hubble parameter, $H_{\rm loc}$. We find that the probability distribution 
of $H_{\rm loc}$ depends not only on the peculiar velocity field resulting from inhomogeneities in the matter distribution 
of a given cosmological model, but also on the distribution of observers used as a prior for the calculation, 
and on the reference frame of the measurement.

For observers randomly distributed in space, the local Hubble parameter is preferentially larger 
than the global expansion rate. This happens due to an uneven volume distribution of voids and overdense regions: 
voids occupy more space and thus they enhance the contribution from cosmic outflows. The excess  
of the local Hubble parameter with respect to the global value, $H_{0}$, 
is stronger when locating observers in the centres of voids. The opposite effect occurs if one distributes 
observers in randomly selected DM haloes. Here the measurement is affected by a large-scale infall and thus the local Hubble 
value is preferentially smaller than the global expansion rate. The deviation from the global Hubble flow depends on the 
mass of the haloes occupied by observers, with $H_{\rm loc}$ being smaller when measured with respect to 
more massive haloes. Among all effects considered in the paper, changing the reference frame from the 
CMB rest frame to the local halo rest frame appears to be the largest. The new reference frame acts as 
a phantom infall so that the local Hubble parameter is smaller than what is measured in the CMB rest frame.

The local Hubble parameter converges to the global value at large scales. Within 
radii $r_{\rm max}\gtrsim 150\hMpc$, the distribution of the local Hubble parameter is well-approximated 
by a Gaussian with a dispersion that can be straightforwardly calculated with linear perturbation theory. At these scales, all the effects 
related to the positions of observers in the cosmic web or the reference frame used are negligible and the intrinsic 
scatter in $H_{\rm loc}$ is fully determined by large-scale inhomogeneities, with values of $1$ and $0.3$ per cent 
within $150\hMpc$ and $300\hMpc$, respectively.

The $68.3$ per cent confidence interval of the relative difference between $H_{\rm loc}$ and $H_{0}$, 
as measured by observers located in Local-Group-like 
DM haloes, is $(-2.3,1.8)$ per cent. After accounting for the redshift selection function of SNIa 
with a redshift range $0.023<z<0.10$ and used for determination of the Hubble constant to a percent precision 
\citep{Rie11}, the scatter in $H_{\rm loc}/H_{0}$ becomes $0.8$ per cent. This means that cosmic variance will be a relevant systematic error in 
upcoming SNIa measurements that are planned for achieving a 1 per cent precision by further improving the 
distance calibration for these objects. The only way to reduce the effect of cosmic 
variance in such measurements is to increase the number of supernovae at large distances. This would shift the 
effective distances of supernovae to scales less affected by inhomogeneities.

We compared our theoretical expectations based on cosmological simulations of the current $\Lambda$CDM model 
with observational constraints on the Hubble parameter at different scales. We 
find that the $2.4\sigma$ discrepancy between the determination of the Hubble parameter from SNIa 
\citep[][]{Rie11} and from CMB observations of \textit{Planck} \citep[][]{Ade13} cannot be ascribed to cosmic 
variance, unless one assumes that the Local Group is located close to the centre of one 
of the largest voids permitted by the standard $\Lambda$CDM cosmological model. 
On the other hand, the $3.6\sigma$ discrepancy between the measurements of $H_{\rm loc}$ from 
SNIa with $z<0.025$ and with $0.025<z<0.10$ \citep{Jha07} can be easily explained by the location of the Local Group in 
an underdense region. When taking into account the scatter in $H_{\rm loc}$ due to inhomogeneities, the tension 
between both measurements is reduced to $2.2\sigma$.

\section*{Acknowledgements}
The Dark Cosmology Centre is funded by the Danish National Research Foundation. RW is grateful to Jens Hjorth 
for his comments and critical reading of the manuscript, and Corinne Toulouse-Aastrup for organising the 
DARK Out workshop. The authors thank the anonymous referee for constructive comments that improved our work.

The Jubilee simulation has been performed on the Juropa supercomputer of the J\"ulich Supercomputing Centre (JSC). 
The BigMultidark simulations  have been performed on the SuperMUC  supercomputer  at the Leibniz-Rechenzentrum (LRZ) 
in Munich,  using the computing resources  awarded to  the PRACE  project  number 2012060963.

AK is supported by the {\it Spanish Ministerio de Ciencia e Innovaci\'on} (MICINN) in Spain through the Ram\'{o}n y 
Cajal programme as well as the grants CSD2009-00064 (MultiDark Consolider project), CAM~S2009/ESP-1496 
(ASTROMADRID network) and the {\it Ministerio de Econom\'ia y Competitividad} (MINECO) through grant AYA2012-31101. 
He further thanks Airliner for the last days of August. 
ITI and WW were supported by The Southeast Physics Network (SEPNet) and the Science and Technology Facilities Council 
grant ST/I000976/1. SH acknowledges support by the Deutsche Forschungsgemeinschaft under 
the grant $\mathrm{GO}563/21-1$. GY acknowledges support from MINECO (Spain) under research grants 
AYA2012-31101 and  FPA2012-34694, Consolider Ingenio SyeC CSD2007-0050 and from the Comunidad de Madrid 
under the ASTROMADRID  Pricit project (S2009/ESP-1496).

\bibliography{master}

\end{document}